\documentclass{iau}
\usepackage{graphicx}

\newcommand{\hMpc}{{\ifmmode{h^{-1}{\rm Mpc}}\else{$h^{-1}$Mpc }\fi}}
\newcommand{\hGpc}{{\ifmmode{h^{-1}{\rm Gpc}}\else{$h^{-1}$Gpc }\fi}}
\newcommand{\hmpc}{{\ifmmode{h^{-1}{\rm Mpc}}\else{$h^{-1}$Mpc }\fi}}
\newcommand{\hkpc}{{\ifmmode{h^{-1}{\rm kpc}}\else{$h^{-1}$kpc }\fi}}
\newcommand{\hMsun}{{\ifmmode{h^{-1}{\rm
{M_{\odot}}}}\else{$h^{-1}{\rm{M_{\odot}}}$}\fi}}
\newcommand{\hmsun}{{\ifmmode{h^{-1}{\rm
{M_{\odot}}}}\else{$h^{-1}{\rm{M_{\odot}}}$}\fi}}
\newcommand{\Msun}{{\ifmmode{{\rm {M_{\odot}}}}\else{${\rm{M_{\odot}}}$}\fi}}
\newcommand{\msun}{{\ifmmode{{\rm {M_{\odot}}}}\else{${\rm{M_{\odot}}}$}\fi}}
\newcommand{\lya}{{Lyman$\alpha$~}}
\newcommand{\apj}{ApJ}
\newcommand{\apjs}{ApJS}
\newcommand{\apjl}{ApJL}
\newcommand{\aj}{AJ}
\newcommand{\mnras}{MNRAS}
\newcommand{\mnrassub}{MNRAS accepted}
\newcommand{\aap}{A\&A}
\newcommand{\aaps}{A\&AS}
\newcommand{\araa}{ARA\&A}
\newcommand{\nat}{Nature}
\newcommand{\physrep}{PhR}
\newcommand{\pasp}{PASP}
\newcommand{\pasj}{PASJ} 

\title[The Local Group in the Cosmic Web] 
{The place of the Local Group in the cosmic web}

\author[Jaime E. Forero-Romero \& Roberto Gonz\'alez]   
{Jaime E. Forero-Romero$^1$ \and Roberto Gonz\'alez$^{2,3}$}
\affiliation{
  
  $^1$Departamento de F\'isica, Universidad de los Andes,
  \\ Cra. 1 No. 18A-10, Edificio Ip \\Bogot\'a, Colombia \\ email:
     {\tt je.forero@uniandes.edu.co} \\[\affilskip]

     $^2$ Instituto de Astrof\'{i}sica, Pontificia Universidad
     Cat\'olica de Chile \\ Av. Vicu\~na Mackenna 4860 \\ Santiago, Chile\\[\affilskip]
 
    $^3$ Centro de Astro-Ingenier\'{i}a, Pontificia Universidad
     Cat\'olica de Chile\\ Av. Vicu\~na Mackenna 4860 \\Santiago, Chile\\
     email: {\tt regonzar@astro.puc.cl}\\[\affilskip]
}

\pubyear{2014}
\volume{308}  
\pagerange{119--126}
\jname{The Zeldovich Universe: Genesis and Growth of the Cosmic Web}
\editors{R. van de Weygaert, S. Shandarin, E. Saar, J. Einasto, eds.}
\begin{document}

\maketitle

\begin{abstract}
We use the Bolshoi Simulation to find the most probable location of the
Local Group (LG) in the cosmic web. Our LG simulacra are pairs of
halos with isolation and kinematic properties consistent with
observations. The cosmic web is defined using a tidal tensor
approach. We find that the LG's preferred location is regions with a
dark matter overdensity close to the cosmic average. This makes filaments and
sheets the preferred environment. We also find a strong
alignment between the LG and the cosmic web. The orbital angular
momentum is preferentially perpendicular to the smallest tidal
eigenvector, while the vector connecting the two halos is strongly aligned
along the the smallest tidal eigenvector and perpendicular to the
largest tidal eigenvector; the pair lies and moves along filaments and
sheets. We do not find any evidence for an alignment between the spin
of each halo in the pair and the cosmic web.  

\end{abstract}

\firstsection 
\section{Introduction}

The kinematic configuration of the Local Group (LG) is not common in
the cosmological context provided by the $\Lambda$ Cold Dark Matter
(CDM) model. 

The LG is dominated by the presence of two spiral
galaxies, the Milky Way (MW) and M31. It is relatively isolated of
other massive structures; the next most luminous galaxy is 10
times less massive than M31, with several dwarf galaxies around a
sphere of $\sim 3$Mpc and the closest massive galaxy cluster, the
Virgo Cluster, is 16.5 Mpc away. The velocity vector of M31 has a low tangential
component and is consistent with a head-on collision toward the MW,
and the velocity dispersion of nearby galaxies up to $\sim 8$ Mpc is
relatively low.

These features make LG analogues uncommon in numerical
simulations. Only about of $2\%$ of MW-sized halos reside in a pair
similar to the MW-M31 in a similar environment. Additionally, the
strong alignments of dwarf satellites of the MW and M31 in the form of
polar and planar structures calls asks for a detailed explanation of
the large scale structure environment that can breed such halos
(\cite{2010MNRAS.407.1449G,lganalogues,sat,ForeroRomero2011,
  2013ApJ...767L...5F}).

Here we present results of the study of the large scale environment of
LG analogues in the context of $\Lambda$CDM. We use a cosmological
N-body simulation (Bolshoi) to infer the most probable place of the LG
in the cosmic web and its alignments. A detailed description
of these results can found in \cite{lgweb}.

\section{Finding the cosmic web in numerical simulations}

We use a web finding algorithm based on the tidal tensor computed from
the gravitational potential field computed over a grid. We define the
tensor as:

\begin{equation}
T_{ij} = \frac{\partial^2\phi}{\partial r_{i}\partial r_{j}}, 
\end{equation}
where the index $i=1,2,3$ refers to the three spatial directions in
euclidean space and $\phi$ is a normalized gravitational potential
that satisfies the following Poisson equation $\nabla^2 \phi=\delta$,
where $\delta$ is the dark matter overdensity.

The algorithm finds the eigenvalues of this tensor, 
$\lambda_1>\lambda_2>\lambda_3$, and use them to classify each cell in
the grid as a peak, filament, sheet or void if three, two, one or none
of the eigenvectors is larger than a given threshold $\lambda_{\rm
  th}$. Each eigenvalue has associated to it an eigenvector ($e_{1}$,
$e_{2}$, $e_{3}$) which are the natural basis to define local
directions in the web. Details describing the algorithm can be found
in \cite[Forero-Romero et al. (2009)]{Tweb}. The dark matter density
is interpolated over a grid with cells of size $0.97$ \hMpc and
smoothed with a Gaussian filter with the same spatial scale. We use a
threshold of $\lambda_{th}=0.25$ to define the different cosmic web
environments.

We use the Bolshoi simulation with a $\Lambda$CDM cosmology described
by the parameters $\Omega_{\rm m}=1-\Omega_{\Lambda}=0.27$,
$H_0=70\,\rm km/s/Mpc$, $\sigma_8=0.82$, $n_s=0.95$
(\cite{2011ApJ...740..102K}). The simulation followed the evolution of dark
matter in a $250 \hmpc$ box with spatial resolution of $\approx
1h^{-1}$~kpc and mass resolution of $m_{\rm p}=1.35\times 10^8\ \rm
M_{\odot}$. Halos are identified with the Bound Density Maxima (BDM) algorithm
(\cite{1997astro.ph.12217K}). The data for the halos and the cosmic
web are available through a public database located at
{\tt http://www.cosmosim.org/}. A detailed description of the database
structure was presented by \cite{Riebe2013}.

\section{Local Groups in cosmological simulations}
We construct a sample of MW-M31 pairs at $z\sim 0$ by using multiple
snapshots from the simulation asking for consistency with the
following criteria:

\begin{itemize}
\item Relative distance. The distance between the center of mass of
  each halo in the pair cannot be larger than $1.3$ Mpc.
\item Individual halo mass. Each halo has a mass in the mass range
  $5\times 10^{11}<M_{200c}<5\times 10^{13}\Msun$.  
\item  Isolation. No neighboring halos more massive than either pair
member can be found within $5$Mpc.
\item Isolation from Virgo-like halos. No dark matter halos with mass
  $M_{200c}>1.5\times 10^{14}\Msun$ within $12$Mpc.
\end{itemize}

With these selection criteria we select close to $6\times 10^3$ pairs to build
a General Sample (GS). From the GS we select two sub-samples
according to the tolerance in kinematic constraints. These sub-samples 
are named $2\sigma$ and $3\sigma$, they correspond to a tolerance of
two and three times the observational errors in the radial velocity,
tangential velocity and separation. The number of pairs in each sample
is $46$ and $120$, respectively.  

\section{The place of the Local Group in the Cosmic Web}

\begin{figure}[t]
\begin{center}
 \includegraphics[width=3.5in]{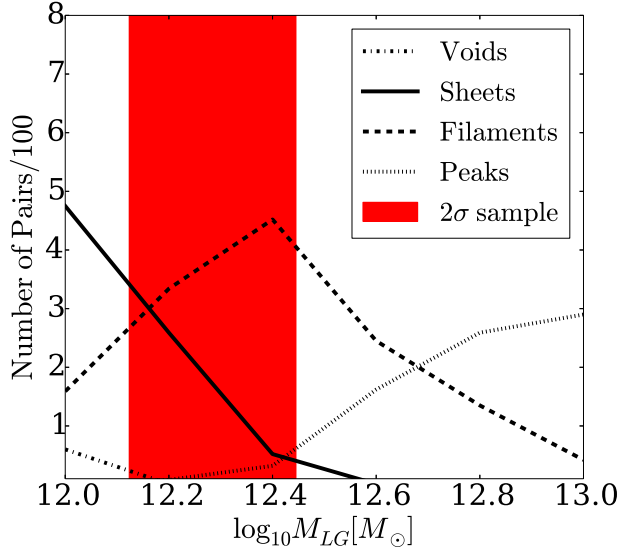} 
 \caption{Number of pairs in a preferred environment as a function of
   the LG total mass. The lines show the results for the General
   Sample, the shaded region shows the mass interval spanned by the
   $2\sigma$  sample. The preferred environments are
   filaments and sheets.}
   \label{fig:environment}
\end{center}
\end{figure}

\begin{figure}[t]
\begin{center}
 \includegraphics[width=3.5in]{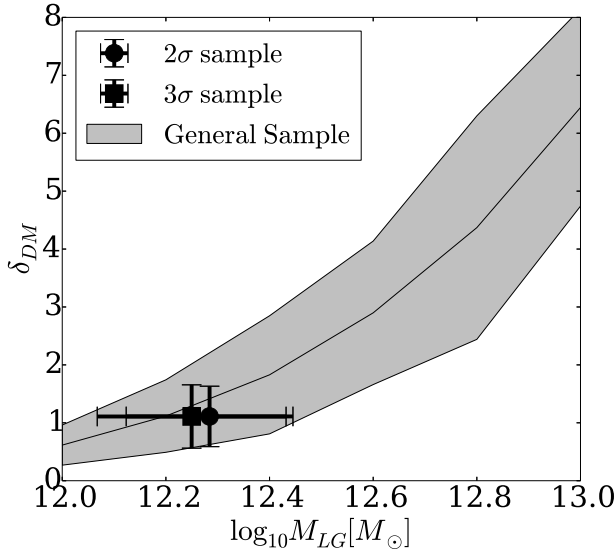} 
 \caption{Dark matter overdensity as a function of the LG total
   mass. The shaded region show the results for the General
   Sample. There is a strong correlation where more massive pairs sit
   in  denser regions. The symbols represent the $2\sigma$ and
   $3\sigma$ samples. The preferred overdensity is in the interval
   $0<\delta<2$.}
   \label{fig:overdensity}
\end{center}
\end{figure}

\begin{figure}[t]
\begin{center}
 \includegraphics[width=3.5in]{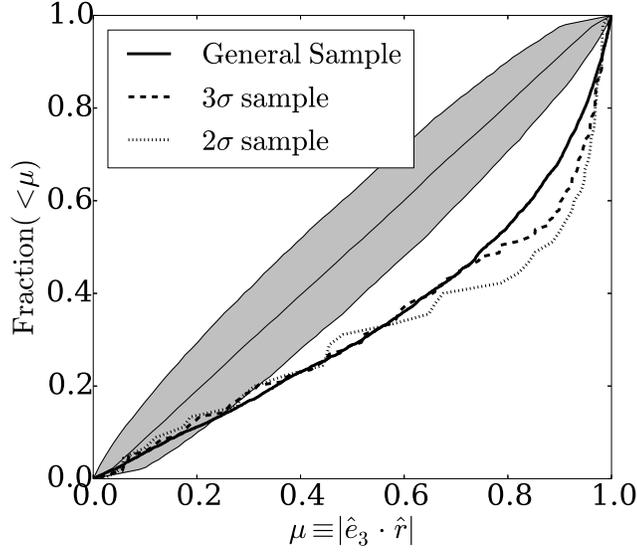} 
 \caption{Cumulative distribution for $\mu=|\hat{e}_{3}\cdot\hat{r}|$
   showing the alignment of the vector joining the two halos in the LG
 with the smallest tidal eigenvector. The shaded regions represents
 the expected result from a distribution of vectors randomly placed in
 space and the corresponding $5\%$ and $75\%$ percentiles for
 distributions with the same number of points as the $2\sigma$ sample.}
   \label{fig:alignment_r}
\end{center}
\end{figure}

\begin{figure}[b]
\begin{center}
 \includegraphics[width=3.5in]{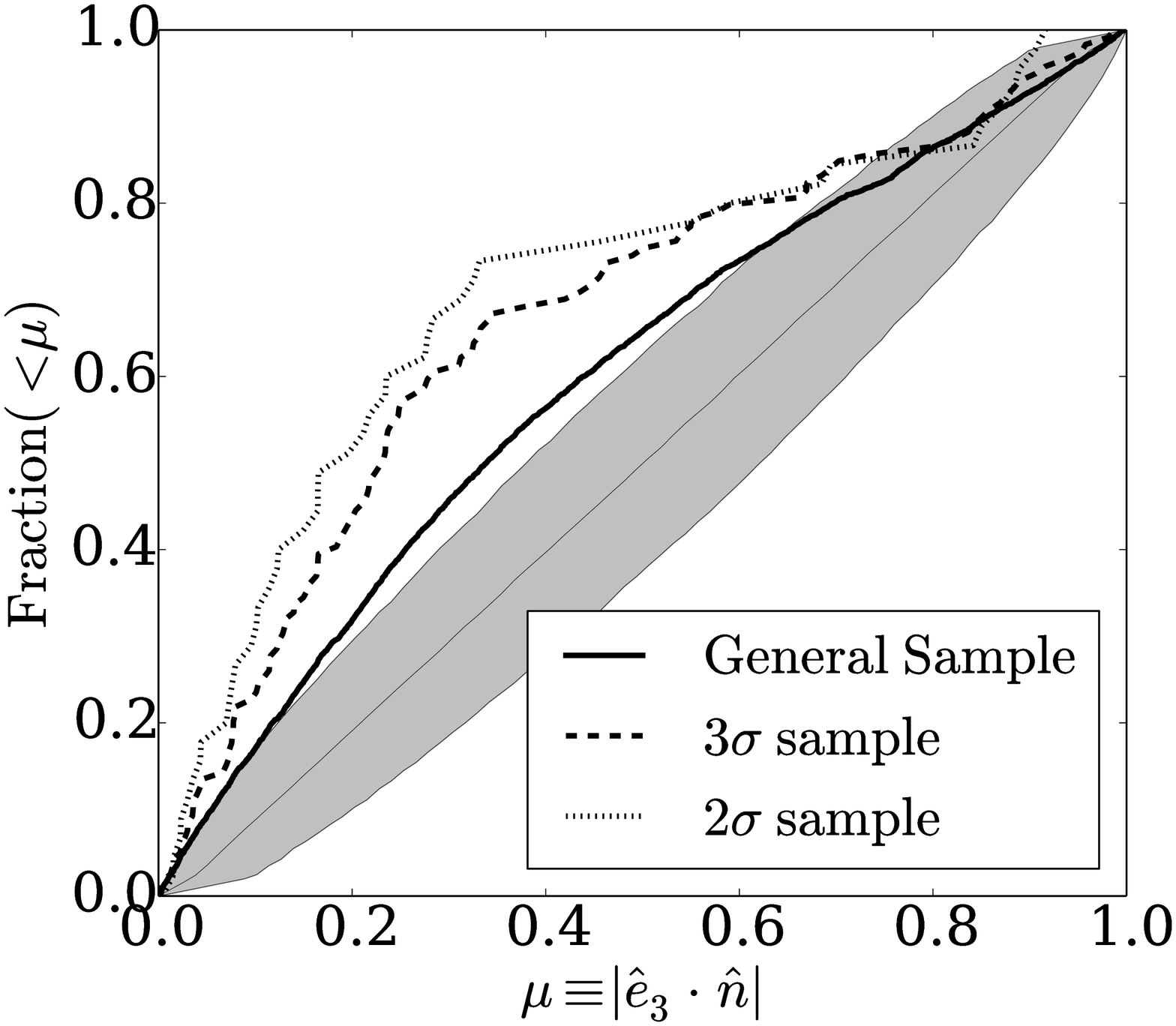} 
 \caption{Same as Figure \ref{fig:alignment_r} for the vector $\hat
   n$ that indicates the direction of the orbital angular momentum of
   the pair.}
   \label{fig:alignment_n}
\end{center}
\end{figure}

We find that the pairs in the general sample are located across
different environments with a strong dependence on  the total pair
mass.  Figure \ref{fig:environment} summarizes this correlation between
environment and mass. Each line represents the 
distribution of pairs in the four different environments. High mass
pairs are mostly located in peaks and filaments while less massive
ones in voids and sheets.  The shaded regions represent the $68\%$
confidence intervals of the mass distributions in the $2\sigma$ 
sample. By and large the LGs in the $2\sigma$ and $3\sigma$ samples are
located in filaments and sheets. In both samples, $\sim 50\%$ of the
pairs can be found in filaments while $\sim 40\%$ are in sheets. 

We can also characterize the preferred place of the pair samples in terms of the
web overdensity.  Figure \ref{fig:overdensity} shows the values of the
overdensity as a function of the total pair mass. The GS and its
uncertainties are represented by the shaded region. The symbols represent the results for the $2\sigma$ and $3\sigma$ samples.   
We see that the range of values for the restricted samples
samples are completely expected from the mass constraint alone. Higher
mass pairs are located in high density regions. The $2\sigma$ and
$3\sigma$ samples have narrower mass range and are located within a
narrower range of overdensities $0.0<\delta<2.0$ peaking at $\delta \sim 1$.

\section{Alignments with the cosmic web}

There is wide evidence showing that DM halo formation properties only
depend on environment through the local DM density. Whether they are
located in sheet or a filament is irrelevant as long the local density
is the same. 

However there is long history of alignment measurement (shape, spin,
peculiar velocities) of individual halos with the cosmic web. In a
recent paper \cite{ForeroRomero2014} presented a study
using the same simulation and cosmic web definition we use here.
They also presented a comprehensive review of all the previous results
from simulations that also inspected the alignment of halo shape and
spin. We refer the reader to the paper for a complete list of
references. 

The main results from the study in \cite{ForeroRomero2014} is that the
halo shape presents the strongest alignment signal. In this case the
DM halo major axis lies along the smallest eigenvector $e_{3}$,
regardless of the web environment. This alignments is stronger for
higher halo masses. Concerning spin alignment the simulations show a
weak anti-alignment with respect to $e_{3}$ for halo masses larger
that $10^{12}$\Msun, and no alignment signal for masses below that
threshold. The peculiar velocities show a strong alignment signal
along $e_{3}$ for all masses.

In our case we test for the alignment of the vector connecting the two
halos ($\hat{r}$), the orbital angular momentum of the pair
($\hat{n}$) and the spin of each
halo. We quantify the alignment using the absolute value of the cosine of
the angle between two vectors $\mu=|\hat{e}_3 \cdot \hat{n}|$ or
$\mu=|\hat{e}_3\cdot \hat{r}|$.

The results for the first two alignments are summarized in Figures
\ref{fig:alignment_r} and \ref{fig:alignment_n}.  We find that the
vector $\hat{r}$ is strongly aligned with $\hat{e}_{3}$, along
filaments and sheets. This trend is already present in the GS and gets
stronger for the $2\sigma$ and $3\sigma$ samples. Concerning the
vector $\hat{n}$ we have a strong anti-correlation  with $\hat{e}_3$,
again this tendency is stronger for the $2\sigma$ and $3\sigma$
samples.

\section{Conclusions}

Here, we have presented results on the expected place of the Local
Group in the cosmic web. Our results are based on cosmological N-body
simulations and the tidal web method to define the cosmic web. We
constructed different Local Groups samples from dark matter halo pairs
that fulfill observational kinematic constraints. 

We found a tight correlation of the LG pairs' total mass with the
local overdensity. For the LG pairs closer to the observational
constraints their total mass is in the range $1\times 10^{12}\Msun <
M_{LG} < 4\times 10^{12}\Msun$ preferred overdensity value is
constrained to be in the range $0<\delta <2$. This restricts the
preferred environment to be filaments and sheets. 

We also found strong alignments of the pairs with the cosmic web. The
strongest alignment is present for the vector joining the two LG
halos. This vector is aligned with the lowest eigenvector and
anti-aligned with the highest eigenvector. This trend is already
present in wide sample of pairs and becomes stronger as the kinematic
constraints are closer to their observed values.

These results  raise the need to use observations to constraint
the alignments of LG pairs with their cosmic web environment. There
are many algorithms available to reconstruct the DM distribution from
large galaxy surveys to tackle this task.  This would allow a direct
quantification of how common are the LG alignments we have found, 
providing a potential new test for $\Lambda$CDM.

J.E.F-R Acknowledges the IAU and the Local Organizing Committee for
providing the financial support to attend this meeting.


\begin{thebibliography}{}


\bibitem[{{Forero-Romero \& {Gonz{\'a}lez} (2014)}}]{lgweb}
{{Forero-Romero}, J.~E., \& {Gonz{\'a}lez}, R.~E.}, ApJ accepted,
ArXiv:1408.3166 

\bibitem[{{Forero-Romero} {et~al.}(2014)}]{ForeroRomero2014}
{Forero-Romero}, J.~E., {Contreras}, S., \& {Padilla}, N. 2014, \mnras, 443,
1090

\bibitem[{{Forero-Romero} {et~al.}(2013)}]{2013ApJ...767L...5F}
{Forero-Romero}, J.~E., {Hoffman}, Y., {Bustamante}, S., {Gottl{\"o}ber}, S.,
\& {Yepes}, G. 2013, \apjl, 767, L5


\bibitem[{{Forero-Romero} {et~al.}(2011)}]{ForeroRomero2011}
{Forero-Romero}, J.~E., {Hoffman}, Y., {Yepes}, G., {Gottl{\"o}ber}, S.,
{Piontek}, R., {Klypin}, A., \& {Steinmetz}, M. 2011, \mnras, 417, 1434


\bibitem[{{Forero-Romero} {et~al.}(2009)}]{Tweb}
{Forero-Romero}, J.~E., {Hoffman}, Y., {Gottl{\"o}ber}, S., {Klypin}, A., \&
{Yepes}, G. 2009, \mnras, 396, 1815


\bibitem[{{Gonz{\'a}lez} {et~al.}(2013)}]{sat}
{Gonz{\'a}lez}, R.~E., {Kravtsov}, A.~V., \& {Gnedin}, N.~Y. 2013, \apj, 770,
96

\bibitem[{{Gonz{\'a}lez} {et~al.}(2014)}]{lganalogues}
---, 2014, \apj, 793, 91


\bibitem[{{Gonz{\'a}lez} \& {Padilla}(2010)}]{2010MNRAS.407.1449G}
{Gonz{\'a}lez}, R.~E., \& {Padilla}, N.~D. 2010, \mnras, 407, 1449


\bibitem[{{Hoffman} et~al.}]{Vweb} {Hoffman} Y., {Metuki} O., {Yepes}
  G., {Gottl{\"o}ber} S., {Forero-Romero} J.~E., {Libeskind} N.~I.,
  {Knebe} A., 2012, \mnras, 425, 2049 

\bibitem[{{Klypin} {et~al.}(2011)}]{2011ApJ...740..102K}
{Klypin}, A.~A., {Trujillo-Gomez}, S., \& {Primack}, J. 2011, \apj,
740, 102

\bibitem[{{Klypin} \& {Holtzman}(1997)}]{1997astro.ph.12217K}
{Klypin}, A., \& {Holtzman}, J. 1997, ArXiv:9712217

\bibitem[{{Riebe} {et~al.}(2013)}]{Riebe2013}
{Riebe}, K., {Partl}, A.~M., {Enke}, H., {Forero-Romero}, J., {Gottl{\"o}ber},
  S., {Klypin}, A., {Lemson}, G., {Prada}, F., {Primack}, J.~R., {Steinmetz},
  M., \& {Turchaninov}, V. 2013, Astronomische Nachrichten, 334, 691



\end{thebibliography}
\end{document}